# Weak antilocalization and topological edge states in PdSn$_4$

N. K. Karn[1,2], M. M. Sharma[1,2] & V.P.S. Awana[1,2,*]

[1]*Academy of Scientific & Innovative Research (AcSIR), Ghaziabad, 201002, India*

[2]*CSIR- National Physical Laboratory, New Delhi, 110012, India*

**Abstract:**

Here we report, successful synthesis of single crystals of topological semimetal (TSM) candidate, PdSn$_4$ using self-flux route. The synthesized crystal is well characterized through X-ray diffraction (XRD), field emission scanning electron microscopy (FESEM), and X-ray photoelectron spectroscopy (XPS). Detailed Rietveld analysis of the powder XRD pattern of PdSn$_4$ confirmed the same to crystallize in Aea2 space group instead of reported Ccce. A large magnetoresistance (MR) along with Shubnikov–de Haas (SdH) oscillations have been observed in magneto transport measurements at 2 K. The presence of weak antilocalization (WAL) effect in synthesized PdSn$_4$ crystal is confirmed and analysed using Hikami Larkin Nagaoka (HLN) formalism, being applied on magnetoconductivity of the same at low magnetic field. An extended Kohler's rule is implemented on MR data, to determine the role of scattering processes and temperature dependent carrier density on transport phenomenon in PdSn$_4$. Further, the non-trivial band topology and presence of edge states are shown through density functional theory (DFT) based theoretical calculations. All calculations are performed considering the Aea2 space group symmetry. The calculated Z2 invariants (0; 0 1 0) suggest the presence of weak topological insulating properties in PdSn$_4$. A clear evidence of topological edge states at Γ point is visible in calculated edge states spectra. This is the first report on PdSn$_4$, showing the presence of SdH oscillation in magneto transport measurements.

**Keywords:** topological semimetal, magneto resistance, weak antilocalization, bulk electronic band structure.

PACS: 74.70.Dd, 74.62.Fj, 74.25.F-

*Corresponding Author
Dr. V. P. S. Awana: E-mail: awana@nplindia.org
Ph. +91-11-45609357, Fax-+91-11-45609310
Homepage: awanavps.webs.com

**Introduction:**

Materials with symmetry protected topological states have recently gathered a significant interest of condensed matter scientists. This class of materials includes topological insulators (TIs), topological semimetals (TSMs), topological superconductors (TSCs) [1-4]. Three dimensional TSMs are the newly discovered quantum materials which include Dirac semimetals (DSMs), Weyl semimetals (WSMs) and nodal line semimetals (NLSMs) [3,4]. Among these, DSMs and WSMs exhibit linear dispersion of bands along all momentum directions. These linearly dispersed bands result in forming of Dirac and Weyl point in bulk electronic band structure in DSMs and WSMs respectively [5]. Apart from DSMs and WSMs, a new class of TSMs is emerged very recently and known as nodal line semimetals (NLSMs) [6]. In contrast of DSMs, NLSMs shows an extended Dirac node along a loop in k-space [7]. Nodal lines in NLSMs can be protected by different symmetries and accordingly these are characterized by different topological invariants. The nodal lines protected by mirror reflection symmetry are characterized by Z invariants, while the ones, which are protected by time reversal symmetry (TRS), inversion symmetry or spin rotation symmetry are characterized by Z2-invariants [8-11].

More recently, some of the homologous stannides viz. $PtSn_4$, $AuSn_4$ and $PdSn_4$ are found to show topological non trivial behaviour and are considered to be the examples of TSMs [11-17]. $PtSn_4$ and $PdSn_4$ both show Dirac node like feature in electronic band structure with topological edge states [12,14,18]. Also, both of these materials are reported to show a high MR % in magneto transport measurements [13,14,18]. The origin of high MR has been a matter of debate and several mechanisms are been proposed for the same, including a perfect electron-hole compensation [19], change in fermi surface structure due to magnetic field [20], and mixing of contributing orbitals [21] etc. Interestingly, the observed high MR in $PdSn_4$ is neither originated from electron hole compensation nor from the Dirac node like feature being observed in electronic band structure [14]. Also, $PdSn_4$ is reported to have two different types of MR % in two available detailed reports on the same [13,14]. $PdSn_4$ is shown to exhibit a quadratic MR % of the order of $10^4$ % having a classical origin [14], while a linear MR (LMR) is also reported [13] in the same upto a magnetic field of 55 T, possibly induced from quantum effects. Besides the magnitude and type of MR [13,14], even the crystallisation space group of $PdSn_4$ is yet debated. Keeping in view of the above facts, it is thus desired to study the $PdSn_4$ in detail.

Here, we report synthesis of single crystals of $PdSn_4$, through a self-flux growth method, which is relatively easier than the earlier reported ones [13,14]. LMR of around 1200

% along with SdH oscillations have been observed in magneto transport measurements, similar to ref. 13. SdH oscillations confirms the presence of π Berry's phase in synthesized $PdSn_4$ single crystal. Kohler scaling law is also applied on magneto transport data of $PdSn_4$, which suggested the presence of two scattering processes. The HLN fitted low field (±0.5 T) magneto conductivity (MC) determines the presence of WAL effect in as grown $PdSn_4$ single crystals. Theoretical investigations are also performed on topological properties of $PdSn_4$ which included the calculation of electronic band structure, Z2 invariants and surface state spectrum. The implemented theoretical calculations suggest the $PdSn_4$ to be a weak topological insulator with strong evidences of topological edge states.

**Experimental & Theoretical Methodology:**

Single crystals of $PdSn_4$ are synthesized using a simple self-flux solid state reactions route as described earlier in case of $AuSn_4$ [16] by us. Contrary to previous reports [13, 14], this method does not require any excess amount of Sn to be added. Synthesized single crystals are silvery shiny and easily cleavable along the growth axis. Rigaku mini flex II table top XRD equipped with Cu-$K_\alpha$ radiation of wavelength 1.54 Å, is used to record XRD pattern on mechanically cleaved crystal flake and gently crushed powder. Here, the XRD pattern is recorded using a slow scan rate of 0.1°/min in order to get a high quality XRD pattern. FESEM image and EDS (Energy Dispersive X-ray Spectroscopy) spectra are recorded using Zeiss EVO-50 made FESEM. XPS measurements are performed using PHI-5000 Versa Probe III made X-ray photoelectron spectrometer. Al-$K_\alpha$ radiation is used to analyse peaks of C 1s peak (reference peak) and the peaks of constituent elements of $PdSn_4$ viz. Sn 3d & Pd 3d. Transport measurements have been carried out on Quantum Design made Physical Property Measurement System (QD-PPMS) equipped with superconducting magnet of capacity ±14 T. A standard four probe method is used to perform magneto transport measurements. Full Prof software is used for Rietveld refinement of powder XRD (PXRD) pattern and VESTA software is used to draw unit cell of the $PdSn_4$ single crystal, which utilize crystallographic information file (CIF) generated from Rietveld refinement.

Computational study of the synthesized material $PdSn_4$, has been performed using the Density Functional Theory (DFT). The DFT based calculations are executed in Quantum Espresso [22,23]. Perdew-Burke-Ernzerhof (PBE) pseudo-potentials are used to incorporate corrections for exchange correlation potentials. The bulk electronic band structure and projected density of states (PDOS) are calculated for simple case as well as including the spin orbit coupling (SOC). For SOC band structure, full relativistic corrected pseudo-potentials are used from the PSEUDOJOJO library. The calculation parameters are as follows: total energy convergence threshold is set to $1.36\times10^{-4}$ eV/atom, kinetic energy cut-off and wave function cut-off are set to 480 Ry and 60 Ry respectively. The first brillouin zone is sampled on a 5×6×6 mesh generated by method given by Monkhrost-Pack [24]. For further analysis, we wannierize the Bloch wavefunctions in the WANNIER90 software [25]. To calculate topological properties, edge states are implemented in Wannier Tools [26]. The model tight-binding hamiltonian is constructed from WANNIER90. The wannier charge centers (WCC) obtained from WANNIER90 are evolved by changing the k-vector. The Z2 invariants are computed by counting the total number of odd or even exchanges of WCC. The surface state spectral function is calculated using the iterative surface Green's function method [27,28] along the plane (110) as implemented in Wannier Tools.

**Results and Discussion:**

Phase purity of synthesized $PdSn_4$ single crystal is verified through Rietveld refinement of PXRD pattern as shown in fig. 1(a). Interestingly $PdSn_4$ is reported to have two space group symmetries viz. Ccce (68) and Aea2 (41). Most of the experimental reports on $PdSn_4$ show the XRD pattern taken on flakes of $PdSn_4$ single crystal, which is not sufficient to determine the space group symmetry of $PdSn_4$ [13,29]. PXRD pattern is shown in ref. 14, in which the XRD peaks are though matched with the Bragg's position of Ccce space group symmetry [30], the respective intensities are not fitted. Another report, which is based on structural refinement of $PdSn_4$ [31] suggested, the same to be crystallized with Aea2 space group symmetry. $PdSn_4$ have similar Bragg positions for the space groups Ccce and Aea2, but the relative intensities of PXRD peaks for both are much different. Here, we performed Rietveld refinement of obtained PXRD pattern of crushed powder of synthesized $PdSn_4$ crystal by considering both of the space group symmetries viz. Ccce and Aea2 and the results are shown in fig. 1(a). The upper panel shows the Rieltveld refined PXRD with Ccce space group and lower one is showing the same with Aea2 space group. As stated earlier, both of the space groups have similar Braggs positions, and XRD peaks are well matched within the applied parameters. It is being observed that the peak intensities, which are considered as the key to determine the space group of $PdSn_4$, are better fitted with Aea2 space group symmetry than as compared to Ccce one. Also, the parameter of goodness of fit i.e., $\chi^2$ is found to be 2.94 for Aea2 space group, which is lower than 6.86 as observed for Ccce space group. Our refinement results show that the synthesized $PdSn_4$ crystal crystallizes within the Aea2 space group symmetry. These results are in agreement with ref. 31, which show $PdSn_4$ to crystallize in Aea2 space group symmetry similar to its homologous compound AuSn$_4$ [15,16,31]. The obtained cell parameters along with the refinement parameters of Rietveld refinement for both of the space groups are given in table 1. CIFs generated from Rietveld refinements for both the space group symmetries are considered to draw the unit cells of synthesized PdSn$_4$ single crystal. Fig. 1(b) and (c) are showing the unit cell structure of PdSn$_4$ in Aea2 and Ccce space groups respectively. Both of the unit cells contain two blocks of PdSn$_4$. The unit cell in Ccce space group is centrosymmetric and the same with Aea2 space group is non-centrosymmetric. Here, Aea2 space group symmetry is considered for further calculations on PdSn$_4$ system.

Unidirectional growth of synthesized PdSn$_4$ crystal is verified through XRD pattern taken mechanically cleaved crystal flake as shown in fig. 1(d). High intensity peaks are observed only for the planes which correspond to the growth axis. The growth axis of PdSn$_4$ is different for both the space groups as the same grows along b-axis for Ccce space group [13,14] and along c-axis for Aea2 space group. Here, Rietveld refined PXRD pattern show the space group symmetry of PdSn$_4$ to be Aea2, which suggest c-axis to be the growth axis of synthesized crystal. The high intensity peaks shown in fig. 1(d) corresponds to reflections due to (002n) planes of PdSn$_4$, which are marked against their respective peaks. These high intensity peaks along specific 2θ angle confirms crystalline character and unidirectional growth of PdSn$_4$ crystal. The reflections due to (002n) planes in XRD pattern are in agreement with the unit cell structure as shown in fig. 1(d).

Crystalline character of synthesized PdSn$_4$ single crystal is also verified by visualizing surface morphology through FESEM image. The left-hand panel of fig. 2 shows the FESEM image of as grown PdSn$_4$ single crystal, depicting a typical terrace type morphology. The observed terrace type morphology suggests layer by layer growth of PdSn$_4$, thus confirming single crystalline character of the same. Also, no colour contrast has been observed in FESEM image, which shows that the synthesized compound is crystallized in single phase. Elemental composition is checked through EDS spectra and the same is shown in right hand panel of fig. 2. EDS spectra shows the peaks corresponding to Sn and Pd without any foreign impurity peaks, confirming the purity of the synthesized crystal. Elemental composition of the constituent elements viz. Sn and Pd is show in inset of right-hand panel of fig. 2. Sn and Pd are found to be 81.2% and 18.8% respectively, suggesting the stoichiometry to be PdSn$_{4.06}$, which nearly matches with the required stoichiometry ratio i.e., PdSn$_4$. FESEM image along with the EDS analysis confirms the phase purity and single crystalline nature of synthesized PdSn$_4$ crystal.

Fig. 3(a) and 3(b) show XPS peaks of PdSn$_4$ in Sn 3d and Pd 3d regions respectively. XPS peaks are calibrated with C 1s peak position, which is taken as reference peak. XPS spectra is both regions are fitted with Gaussian fitting formula. Fig. 3(a) represents XPS peaks in Sn 3d regions, which shows the presence of four peaks at 484.41±0.006 eV, 486.52±0.005 eV, 492.83±0.009 eV and 494.93±0.007 eV. The observed peaks at 484.41±0.006 eV and 492.83±0.009 eV correspond to peaks due to spin orbit doublet of Sn viz. Sn 3d$_{5/2}$ and Sn 3d$_{3/2}$. These peaks are separated by 8.42 eV, which closely matches with the standard value of 8.4 eV [32], also these values are in agreement with the previous report on XPS measurement of PdSn$_4$ [29]. The other two peaks observed in Sn 3d regions are attributed to the peaks of SnO, which occurred due to surface oxidation of the sample due to air exposure. It has been reported [29], that a slight air exposure of PdSn$_4$ crystal leads to surface oxidation. Fig. 3(b) represents XPS peaks of PdSn$_4$ single crystal in Pd 3d region, which shows the presence of only two peaks of spin orbit doublets of Pd 3d orbital viz. 3d$_{5/2}$ and 3d$_{3/2}$. Interestingly, no peak for palladium oxide is visible in XPS spectra as observed in Sn 3d region. This result is in well agreement with ref. 32, which also shows that the XPS spectra in Pd 3d region remains intact with air exposure. Less electronegativity of Sn atoms as compared to Pd atoms, makes Sn more prone to lose electrons as compared to Pd, which results in formation of SnO phase at the surface of crystal. XPS peaks of spin orbit doublet of Pd 3d core shell are observed at 335.98±0.004 eV and 341.26±0.003 eV, these peaks are attributed to Pd 3d$_{5/2}$ and Pd 3d$_{3/2}$. The observed binding energies of spin orbit doublet of Pd 3d core shell are in agreement with ref. 29. These peaks are slightly shifted from the standard position of XPS peaks for elemental Pd 3d core shells [32], which occurs due to bonding between Sn and Pd atoms. The separation between XPS peaks in Pd 3d region is found to be 5.28 eV, which closely matches with the standard value of 5.26 eV [32]. All XPS peaks of PdSn$_4$ in Sn 3d and Pd 3d region are listed in table 2 with their respective binding energy and full width at half maxima (FWHM).

Fig. 4 represents resistivity vs temperature (ρ-T) measurements of synthesized PdSn$_4$ single crystal at zero magnetic field. Resistivity tends to decrease sharply with temperature, showing high metallicity of the synthesized crystal. The experimental ρ-T data is fitted in whole temperature range i.e., 2 K-300 K using Bloch-Grüneisen formula, which shows that ρ(T) can be well described by the following formalism;

$$\rho(T) = \left[\frac{1}{\rho_s} + \frac{1}{\rho_i(T)}\right]^{-1} \qquad (1)$$

Here, ρ$_s$ denotes the temperature independent saturation resistivity and ρ$_i$(T) is the temperature dependent term and can be given by following equation,

$$\rho_i(T) = \rho(0) + \rho_{el\text{-}ph}(T) \tag{2}$$

here ρ(0) denotes residual resistivity arising due to impurity scattering and the second term $\rho_{e\text{-}ph}(T)$ denotes the temperature dependent term, which depends on electron-phonon scattering. Further, $\rho_{e\text{-}ph}(T)$ is given by the following formula

$$\rho_{el-ph} = \alpha_{el-ph} \left(\frac{T}{\theta_D}\right)^n \int_0^{\frac{\theta_D}{T}} \frac{x^n}{(1-e^{-x})*(e^x-1)} dx \tag{3}$$

here $\alpha_{el\text{-}ph}$ is electron-phonon coupling parameter, $\theta_D$ represents Debye temperature and n is constant. ρ-T data is well fitted with the above equation for n=5, signifying dominant electron-phonon scattering. Residual resistivity ratio (RRR) is found to be 27, which shows high metallicity of the synthesized crystal. The observed value of RRR is lower than the previously reported values for PdSn$_4$ crystal [13,14,29], still it is sufficiently high to show high metallicity present in the synthesized crystal. The obtained value of Debye temperature $\theta_D$ from above fitting formula is 177 K.

To study about the topological properties of synthesized PdSn$_4$ single crystal, magneto transport measurements have been carried out in magnetic field range of ±12 T. Fig. 4(b) represents MR % vs H plot for applied field upto ±12 T at different temperatures viz. 2 K, 5 K, 10 K, 20 K, 50 K and 100 K. MR % has been calculated using the following formula

$$MR\ \% = \frac{\rho(H)-\rho(0)}{\rho(0)} \times 100 \tag{4}$$

Here ρ(H) denotes the resistivity in presence of applied magnetic field, while ρ(0) denotes the zero-field resistivity. The synthesized PdSn$_4$ single crystal is found to show a non-saturating high MR % of around 1200 % in field range of ±12 T at 2 K. The observed MR is found to have linear dependency on the applied magnetic field, this result is in agreement with ref. 13. The MR % tends to decrease as the temperature is increased and the same is found to be around 20 % at 100K. The shape of MR % plot provides important information about the conduction mechanism, as for conventional metals MR shows quadratic dependency over the magnetic field and this behaviour is well explained by classical theories. In contrast, most of the TSMs exhibit non saturating LMR originated due to quantum effects [33-35]. LMR can also be observed due to inhomogeneity of the sample [36], but this case is neglected for the synthesized PdSn$_4$ crystal as the same is found to show high metallicity in ρ-T measurements with a quite high RRR. Interestingly, PdSn$_4$ is shown to have quadratic MR in ref. 14 similar to conventional metals. Our results are in good agreement with ref. 13 showing the presence of LMR in PdSn$_4$ upto a magnetic field of 55 T.

PtSn$_4$, the homologous compound of PdSn$_4$ is reported to show SdH oscillations [18], in magneto transport measurements. Here we examined the magneto transport measurement of PdSn$_4$ at 2 K, to check whether the SdH oscillations are present in the system or not. In fig. 4(c), dρ/dH is plotted against the inverse magnetic field (H$^{-1}$) for PdSn$_4$ at 2 K. In fig. 4(c), dρ/dH depicts oscillatory motion with successive maxima and minima, this shows that the SdH oscillation are present in measured PdSn$_4$ single crystal. The SdH oscillations are important in order to get information about the different Fermi surfaces inside the Brillouin zone. The frequency of SdH oscillation determines the cross section area of Fermi surface. Here, in order to determine the frequency of SdH oscillations, Fast Fourier transform (FFT) of SdH oscillations has been performed, and the same is shown as the inset of fig. 4(c). Three major peaks are observed in FFT plot, these are labelled as α, 2α and β with corresponding frequencies 51.10, 102.20 and 136.55 T. These frequencies are similar to as observed from SdH oscillations in PtSn$_4$ [18], which is a homologous compound of PdSn$_4$. The cross section area of Fermi surfaces corresponding to different frequencies has been calculated using the Onsager relation, $F = \left(\frac{\hbar}{2\pi e}\right) A_F$. Here $\hbar$ and e are constants. The obtained values of cross section area at different frequencies are given in table 3. Here, we considered the Fermi surface to have circular cross section, and the fermi wave vector is calculated using the formula $k_F = \left(\frac{A_F}{\pi}\right)^{1/2}$. The obtained values of k$_F$ corresponding to different frequencies are given in table 3. Fermi wave vector is related to the 2D surface carrier density (n$_{2D}$) with the relation, $n_{2D} = \frac{k_F^2}{4\pi}$. The obtained values of n$_{2D}$ at different frequencies are also given in table 3.

SdH oscillations are also important in investigation of presence of π Berry phase in topological materials using the Landau level (LL) fan diagram. LL fan diagram is shown in fig. 4(d), in which the Landau index n has been plotted against the inverse magnetic field. LL index has been assigned to the maxima of quantum oscillations as given in main panel of fig. 4(c). LL index follows the Lifshitz-Onsager quantization rule, according to which the relationship between n and applied magnetic field can be expressed as $\frac{A_F \hbar}{eH} = 2\pi \left(n + \frac{1}{2} + \beta\right)$, where 2πβ is Berry's phase. In order to determine the Berry's phase, the LL fan diagram is linearly fitted, as shown in fig. 4(d). The intercept of linearly fitted LL fan diagram determines the Berry's phase in the studied system. Here, the obtained value of intercept is found to be -0.526. This value of intercept corresponds to 1.052π Berry's phase. This value of Berry's phase is very close to π, which confirms that the observed SdH oscillations are originated due to topological edge states [37,38]. This also shows that the conduction mechanism at 2 K is governed by 2D topological edge states.

Another evidence of the presence of topological edge states can be found in low field MR % data, in which a V type cusp has been observed as shown in inset of fig. 4(b). This V type cusp in low field MR data occurs due to the presence of weak localization (WAL) effect [39]. In topological materials, WAL arises due to delocalization of electrons caused by destructive interference between the two-time reversed paths of electron, which are associated with π Berry phase. This π Berry phase is disturbed by application of magnetic field and causes appearance of positive MR with V type cusp at low magnetic fields. Here, also the V type cusp is observed in low field MR data, which tends to be broadened as the temperature is increased. The broadening of V type cusp at higher temperature is associated with suppression of WAL effect at higher temperatures. The suppression of WAL effect at higher temperatures also signifies that the contribution of topological edge states becomes weaker at higher temperatures.

Kohler scaling law has been applied on magneto transport data to determine the contribution of various scattering mechanism in transport phenomenon in synthesized $PdSn_4$ single crystal. In ref. 14, $PdSn_4$ is shown to follow Kohler scaling law, due to which MR % was shown to have classical origin, which was in agreement with the observed quadratic MR in the same report. Here, the observed MR has linear dependency on applied field at low temperature and it becomes important to apply Kohler scaling law to check whether the scattering process remains same as the temperature is increased or not. According to Kohler's rule, magneto resistance can be described in terms of Hτ, where τ is scattering time and defined as the time spent between two successive scattering events and inversely proportional to zero field resistivity. Here, an increment in H leads to shrinking of electronic orbitals which further leads to decrement in scattering time τ, thus making the product Hτ a constant value. For the materials, which follows Kohler scaling law the response of magnetoresistance should be invariant with change in Hτ at different temperatures i.e., all the magneto resistance curves should merge into each other. This shows that only single scattering process in present at all temperatures [40-42]. Here, fig. 5(a) shows the variation of MR % with respect to $H/ρ_0$ i.e., Hτ at temperatures viz. 2, 5, 10, 20, 50 and 100 K. Interestingly, the MR % curves do not merge into each other, which is in contrast with ref. 14. Here, we found to that Kohler scaling law is not applicable for synthesized $PdSn_4$ single crystal. Violation of Kohler's rule in synthesized $PdSn_4$ single crystal is in agreement with the other reported TSMs [43-47]. The violation of Kohler's rule shows that there may be more than one scattering process present in synthesized PdSn$_4$ single crystal. The presence of multiple scattering processes will be confirmed in the later part of the MS. Interestingly, all MR% curves are found to be nearly parallel to each other. It indicates that the extended Kohler's rule can be applied to PdSn$_4$. For this, we added an additional multiplier term to H/ρ$_0$, which is 1/n$_T$ and known as temperature dependent carrier density. Now we tried to scale MR% curve with H/(ρ$_0$n$_T$) i.e., MR% is taken as a function of H/(ρ$_0$n$_T$). So, MR% can be represented as $MR \sim f\left(\frac{H}{\rho_o n_T}\right)$. Fig. 5(b) shows the MR% vs H/(ρ$_0$n$_T$) curves at various temperatures viz. 2, 5, 10, 20, 50 and 100 K. Here, all MR% curves are scaled to MR% curve at 100 K, i.e., for 100 K the value of n$_T$ is taken to be 1. The observed values of n$_T$ for other temperatures are found to be 0.052, 0.028, 0.0265, 0.024 and 0.023 for 50, 20, 10, 5 and 2 K respectively. It is clear from fig. 5(a), that the MR% curve at 100 K is more deviated from all other MR% curves as the observed MR% is much more lower in comparison to other temperatures. Similar result is observed in extended Kohler's rule, where the obtained values of n$_T$ for the measured temperatures are much lower than that of at 100 K. Overall, by applying extended Kohler's rule we observed that the all MR% curves are merged into each other. This suggests that the temperature dependent carrier density also plays a role in violation of Kohler's rule in PdSn$_4$. Also, the merged curves are fitted with power law i.e., MR%=AH$^m$, as shown by dotted line in fig. 5(b). The value of exponent m is found to be 0.9, which is near to 1 showing the presence of linear MR in PdSn$_4$.

V-type cusp observed in MR% vs H plot at low magnetic fields provided some hints of presence of WAL effect is PdSn$_4$. Here, the low magnetic field (±0.5 T) MC of synthesized PdSn$_4$ single crystal is analysed using the HLN formalism to get more insight about the observed WAL effect and topological edge states. HLN equation models the difference in MC i.e., Δσ(H), which is given by σ(H)-σ(0). HLN equation used to model the low magnetic field MC data is as follows [48]:

$$\Delta\sigma(H) = \frac{-\alpha e^2}{\pi h}\left[ln\left(\frac{B_\varphi}{H}\right) - \Psi\left(\frac{1}{2} + \frac{B_\varphi}{H}\right)\right] \quad (5)$$

Here, B$_\phi$ represents characteristic field and is given by $B_\varphi = \frac{h}{8e\pi l_\varphi^2}$, and Ψ is digamma function. l$_\phi$ represents the phase coherence length, which is the maximum length travelled by the electron while maintaining its phase. The pre factor α in HLN equation provides important information about the presence of weak localization (WL) or WAL effect. Fig. 6(a) shows the HLN fitted MC data upto ±0.5 T at various temperatures viz. 2 K, 5 K, 10 K, 20 K and 50 K. The values of α are found to be negative for all temperatures, which suggest the presence of WAL effect in PdSn$_4$ single crystal. The observation of WAL effect can have different origins as it can be induced due to intrinsic high SOC of the material or due to the presence topological edge states protected by π Berry phase. In the present case, the value of α at 2 K, is found to be -0.96(2), which is near to standard value i.e., -1 as suggested for the topological materials with two independent edge states [49,50] protected by π Berry phase. The presence of π Berry phase in synthesized PdSn$_4$ single crystal is also confirmed through observed SdH oscillations at 2 K. These facts suggests that the observed WAL effect in PdSn$_4$, is originated due to topological edge states protected by π Berry phase. Also, the observed value of α suggests that the conduction mechanism is solely contributed by topological edge states at 2 K. At higher temperatures, the value of α shifts from -1 suggesting bulk conducting channels starts to contribute in the conduction mechanism. The negative value of α decreases monotonically with temperature as shown by blue curve in fig. 6(b). Further, at 100 K MC cannot be fitted by HLN equation, as it shows quadratic dependency on applied magnetic field. MC at 100 K is fitted by quadratic term $\beta H^2 + c$ [51], and the same is shown in inset of fig. 6(a). Here, the coefficient of quadratic term i.e., β accounts the effects of scattering events and cyclotronic MR and the constant term c symbolises the contribution due to defects present in the measured sample. The MC data at 100 K is found to be well fitted with the mentioned quadratic term, showing that the contribution of topological edge states in conduction mechanism is totally suppressed, and rather the same is fully governed by bulk conducting states. The observed variation of pre factor α can be related to temperature dependent carrier density (n$_T$) as obtained from extended Kohler's rule. The extended Kohler's rule suggested that there is much variation in n$_T$ at all measured temperatures namely, 2, 5, 10, 20 and 50 K in comparison to the same at 100 K. This is related to the fact that at 100 K, transport phenomenon is totally governed by bulk conducting channels resulting into larger carrier density. Instead, the topological edge states are the major contributor to transport phenomenon at other lower temperatures, resulting in lower carrier density. At 2 K, the transport properties are governed by two independent topological edge states, which accounted for the lowest carrier density at 2 K.

It is clear from fitted data at different temperatures, that the magneto response of electrical conductivity in synthesized PdSn$_4$ crystal has different origins. HLN modelling of MC also provides the value of phase coherence length at different temperatures. In fig. 6(b), the variation in $l_\phi^{-2}$ with respect to temperature is shown by red symbols. The variation in $l_\phi$ with respect to temperature provides important insights about the dephasing mechanism and scattering processes. The Nyquist criterion suggests that only electron-electron (e-e) scattering is included in dephasing mechanism if $l_\phi^{-2}$ varies linearly with temperature [52,53], while a deviation of $l_\phi^{-2}$ vs T plot from linearity suggests that there exists another scattering process, which is commonly taken as electron-phonon (e-p) scattering. It is clear from fig. 6(b), that the variation in $l_\phi^{-2}$ is not linear with respect to temperature, suggesting that the dephasing mechanism in synthesized $PdSn_4$ crystal is contributed by two scattering processes viz. e-e scattering and e-p scattering. The presence of two scattering process strengthens our result of violation of Kohler rule in MR vs $H/\rho_0$ plot. Further, the $l_\phi^{-2}$ vs T plot is fitted with the following power law [54] shown by solid black line in fig. 6(b):

$$\frac{1}{l_\Phi^2(T)} = \frac{1}{l_\Phi^2(0)} + A_{e-e}T^p + A_{e-p}T^q \qquad (6)$$

Here $l_\phi(0)$ denotes the dephasing length at absolute zero and $A_{e-e}T^p$ and $A_{e-p}T^q$ denote the contribution from e-e and e-p scattering respectively. The value of p and q from fitting are found to be 1 and 2 respectively, which shows the 2-D nature of observed WAL effect, which is observed in other topological materials [55]. The obtained value of $l_\phi(0)$ from power law fitting is 118 nm, and the values of $A_{e-e}$ and $A_{e-p}$ are found to be $4.452\times10^{-7}$ and $1.08\times10^{-7}$. The values of pre factor $\alpha$ and phase coherence length $l_\phi$ at various temperatures obtained from HLN fitting are listed in table-4.

To understand more about topological properties of $PdSn_4$ system, the bulk electronic band structure along with the DOS is calculated by first principle methods using Quantum Espresso. A converged electron density functional and Bloch wavefunctions are obtained using self-consistent functional calculation with electronic convergence cut-off $10^{-10}$ units. The cell parameters are taken from the XRD pattern analysis of the synthesised material and since the resulting structure belong to space group Aea2, the ibrav was set to 91. The high symmetric path in the first Brillouin zone for the band structure calculation is Z-Γ-S-R-Z-T-Y-Γ and the same is shown in fig. 7(a). The path is taken considering the optimised one [56]. Figure 7(b) and 7(c) show the calculated band structure without SOC (w/o-SOC) and with SOC parameters respectively. All bands are plotted with respect to Fermi level i.e., setting $E_f=0$. We find that there are five Bands, which cross the fermi level. For band structure w/o-SOC parameters, we find several bands crossing each other indicating possibility of Dirac cones. When we include SOC in band structure calculation, all the band degeneracies are removed. Figure 7(d) and 7(f) show the PDOS for w/o-SOC and SOC parameters respectively. The total DOS is finite at the fermi level confirming the metallic/semimetallic behaviour. The projected DOS clearly indicate that near the Fermi level the bands have major contributions from the s and d orbitals of Pd atom and p orbitals of Sn atoms.

The Bloch wavefunction obtained from first principle calculation are wannierized using WANNIER90 software. The wannierization process takes the wavefunction representation in k-space to real space. The maximally localised wannier functions (MLWFs) are obtained with the disentanglement cut-off set to $10^{-10}$ a.u. and the convergence cutoff to $10^{-6}$ a.u. Out of 92 w/o-SOC Bloch wave-functions, only 44 are wannieriezed. Using the wannier functions the band structure is reproduced in the range ±2.5 eV about the Fermi level on the same size of first Brillouin zone mesh of 5×6×6. The fat band analysis in the WANNIER90 confirms the same results of orbital contributions near the fermi level.

Now, we use these wannierised bands to calculate the band dispersion in kz=0 plane implemented in Wannier TOOLS on a 81×81 mesh. Figure 8(a) and (b) shows the band dispersions for without SOC and with SOC parameters. Our calculation shows the band dispersion in the plane for without SOC bands are of Dirac cone type with linear dispersion. But, since they are degenerate within the calculation limit, the associated Fermions are massless. But when we include effective SOC, we find that Dirac cone is gapped out and in place of linear dispersion, parabolic dispersion is present. This parabolic dispersion indicates heavy fermions (non-zero mass) present in the system. This suggest that the system can be described by Dirac Hamiltonian. However, for more detailed and exact analysis, k.p theory based Hamiltonian need to constructed which is beyond scope of our study.

Further, we calculate the edge state spectrum using the iterative green's function implemented in Wannier Tools. The edge state spectrum of $PdSn_4$ has been studied from DFT calculations and experimentally tested by the Angle resolved photoelectron spectroscopy [12, 14]. However, in the mentioned earlier reports [12,14], $PdSn_4$ has been studied by considering Ccce space group. From experimental results of our study, we considered $PdSn_4$ to be crystallized with Aea2 space group symmetry. So, we calculate edge spectrum for the first time by considering the Aea2 space group. The surface card is set to be the plane (110), and the calculation is performed by taking 101 slices of one reciprocal vector. The plane (110) is shown in figure 9(a), and in that plane, the path taken for edge state spectrum calculation is Z → Γ → S. Figure 9(b) and (c) show the obtained edge spectrum and bulk spectrum respectively. A possible Dirac cone is observable at Γ point near energy −0.25 eV in both the spectrum. A similar Dirac cone of type-II is observable very close to Fermi energy again at around -0.25eV in between the path Z → Γ (fig.9b), but not at the high symmetry point in edge state spectrum.

Recently, a series of theoretical developments has allowed us to classify and quantify the topology present in the system by evaluating Z2-invariants. Since, the system $PdSn_4$ with non-centrosymmetric space group, is non-magnetic and respects the time-reversal symmetry. Therefore, to analyse the topology of the electronic bands, we further calculate the Z2 invariant. Different Z2-states are separated by a topological phase transition, which involves an adiabatic transformation of the bands leading to the gap closing of the two bands. Thus, indirectly it indicates any possible band inversion leading to non-trivial topological states. Till now, there are three methods developed to calculate Z2 invariant— Pfaffians/parity calculation over the fixed points of the time reversal symmetry [57], Fukui- Hatsugai method [58] and by counting the windings of WCC evolution around the BZ [59, 60]. Here we adopt WCC method to determine Z2 invariant which has been adjoined with WANNIER TOOLS for numerical implementation. The WCCs are the pseudo charge points. Their locations are the extrema points of probability distribution of MLWFs, locations of maximum probability of MLWFs regarded as negative charge and with least (zero) probability regions are regarded as positive charge. Exchange of these charge centres in different k-planes define trivial and non-trivial topology. So, from the tight binding model obtained for Wannier functions, the WCC are computed numerically and evolved in the 6-planes— $k_x, k_y, k_z = 0$ and $k_x, k_y, k_z = 0.5$ in the Brillouin zone. An odd number of the crossing of WCC implies a topologically nontrivial state (Z2 = 1), whereas an even number of crossings indicate the presence of a topologically trivial state (Z2 = 0). The w/o-SOC bands have degeneracies but the SOC bands are gapped out, which enable us to calculate Z2 invariant with SOC bands. The Z2 invariant has four indices ($\upsilon_0; \upsilon_1 \upsilon_2 \upsilon_3$). From the results obtained in Wannier Tools, we find that first index has some redundancy in $k_x$-$k_z$ plane, while the rest three are (0 1 0), which indicates the presence of weak topologically insulating phase in the present system. For convention, hence the Z2 index for $PdSn_4$ can be given as (0; 0 1 0).

**Conclusion:**

Summarily, $PdSn_4$ single crystals have been synthesized without taking excess Sn amount as flux. Synthesized crystals are well characterized in context of phase purity, morphology and chemical environment around constituent elements. Magneto transport measurements evidenced the presence of non-saturating LMR, which is accompanied by SdH oscillations confirming the presence of π Berry phase in $PdSn_4$. Kohler's rule is found to be violated due to multiple scattering process and the temperature dependent carrier density. The presence of topological edge states induced WAL effect is also evidenced through HLN modelling of low magnetic field MC of as grown $PdSn_4$ crystals. The presence of topological edge states is also evidenced in theoretical calculations on $PdSn_4$ system, which clearly shows the system exhibit weak topology with signatures of topological edge states at around Γ point in edge states spectrum.

**Acknowledgement:**

The authors would like to thank Director NPL for his keen interest and encouragement. Authors would like to thank Dr. J. S. Tawale, CSIR-NPL for FESEM and EDS measurements. Authors would like to thank Ms. Prachi, IIT-Roorkee for XPS measurements. M.M. Sharma and N.K. Karn would like to thank CSIR, India for the research fellowship. M.M. Sharma and N.K. Karn are also thankful to AcSIR for Ph.D. registration.

**Table-1**

Parameters obtained from Rietveld refinement:

| Space group: Aea2 (41) | | Space Group: Ccce (68) | |
|---|---|---|---|
| Cell Parameters | Refinement Parameters | Cell Parameters | Refinement Parameters |
| Cell type: Orthorhombic<br>Space Group: Aea2 (41)<br>Lattice parameters: a=6.393(7)Å<br>b=6.434(7)Å & c=11.449(1) Å<br>α=β=γ=90°<br>Cell volume: 470.926Å$^3$<br>Density: 8.203 g/cm$^3$<br>Atomic co-ordinates:<br>Sn1 (0.1871,0.3275,0.1437)<br>Pd (0,0,0)<br>Sn2 (0.3404,0.1695,0.8871) | $\chi^2$=2.94<br>$R_p$=22.1<br>$R_{wp}$=21.6<br>$R_{exp}$=12.58 | Cell type: Orthorhombic<br>Space Group: Ccce (68)<br>Lattice parameters: a=6.436(3)Å<br>b=11.460(6)Å & c=6.394(6) Å<br>α=β=γ=90°<br>Cell volume: 471.6Å$^3$<br>Density: 18.589 g/cm$^3$<br>Atomic co-ordinates:<br>Sn1 (0.3357,0.1241,0.1542)<br>Pd (0,0.25,0.25) | $\chi^2$=6.86<br>$R_p$=38.3<br>$R_{wp}$=34.8<br>$R_{exp}$=13.29 |

**Table-2**

XPS peaks position and FWHM of constituent elements of synthesized PdSn$_4$ single crystal:

| Element | Spin-orbit doublet | Binding Energy | FWHM |
|---|---|---|---|
| Sn | 3d$_{5/2}$ | 484.41±0.007 eV | 1.00±0.01 eV |
|    | 3d$_{3/2}$ | 492.83±0.010 eV | 0.94±0.02 eV |
| Pd | 3d$_{5/2}$ | 341.26±0.001 eV | 0.82±0.01 eV |
|    | 3d$_{3/2}$ | 335.98±0.004 eV | 0.90±0.01 eV |

**Table: 3**

Parameters obtained from SdH oscillations:

| Frequency | $A_F$ (m$^{-2}$) | $k_F$ (Å) | $n_{2D}$ (cm$^{-2}$) |
|---|---|---|---|
| α (51.10 T) | 4.87×10$^{17}$ | 0.039 | 1.2×10$^{12}$ |
| 2α (102.2 T) | 9.74×10$^{17}$ | 0.055 | 2.4×10$^{12}$ |
| β (136.55 T) | 1.30×10$^{18}$ | 0.064 | 3.2×10$^{12}$ |

**Table: 4**

Low field (up to 1 Tesla) HLN fitted parameters of PdSn$_4$ single crystal

| Temperature(K) | α | $l_\phi$ |
|---|---|---|
| 2 | -0.96(2) | 117.45 nm |
| 5 | -0.82(6) | 112.42 nm |
| 10 | -0.48(8) | 108.42 nm |
| 20 | -0.32(2) | 97.45 nm |
| 50 | -0.23(1) | 57.10 nm |

**Figure Captions:**

**Fig. 1(a)** Rietveld refined PXRD pattern of synthesized PdSn$_4$ single crystal in which upper panel is showing the fitted plot considering the Ccce space group symmetry and the lower panel is showing the same considering the Aea2 space group symmetry **(b)** Unit cell of PdSn$_4$ with Aea2 space group symmetry **(c)** Unit cell of PdSn$_4$ with Ccce space group symmetry **(d)** XRD pattern taken on mechanically cleaved crystal flake of PdSn$_4$.

**Fig. 2** FESEM image and EDS spectra of synthesized PdSn$_4$ single crystal.

**Fig. 3** XPS spectra of synthesized PdSn$_4$ single crystal in **(a)** Sn 3d region **(b)** Pd 3d region.

**Fig. 4 (a)** B-G fitted resistivity vs temperature (ρ-T) plot of synthesized PdSn$_4$ single crystal. **(b)** MR % vs H plot of synthesized PdSn$_4$ single crystal at temperatures viz. 2 K, 5 K, 10 K, 20 K, 50 K & 100 K in a magnetic field range of ±12 T in which inset is showing the power the same at low magnetic field of ±1 T. **(c)** dρ/dH vs H$^{-1}$ plot of PdSn$_4$ at 2 K signifying the presence of SdH oscillations in the same, the inset is showing the FFT results of SdH oscillations having three peaks. (d) Linearly fitted LL fan diagram of PdSn$_4$.

**Fig. 5(a)** MR% vs H/ρ$_0$ plots for Kohler's scaling of MR at various temperatures viz. 2 K, 5 K, 10 K, 20 K, 50 K and 100 K. (b) MR% vs H/(n$_T$ρ$_0$) plots for extended Kohler's scaling of MR at various temperatures viz. 2 K, 5 K, 10 K, 20 K, 50 K and 100 K.

**Fig. 6 (a)** HLN fitted low magnetic field (±0.5 T) MC of synthesized PdSn$_4$ single crystal at temperatures 2 K, 5 K, 10 K, 20 K and 50 K, in which inset is showing the MC at 100 K fitted with equation βH$^2$+c. **(b)** Variation of HLN fitting parameters with respect to temperature.

**Fig. 7(a)** The first Brillouin zone with high symmetric points, the green arrow shows the path chosen for the Band structure calculation. **(b)** bulk electronic band structure calculated without considering SOC parameters **(c)** bulk electronic band structure calculated with SOC parameters. **(d)** The calculated PDOS of PdSn$_4$ system w/o-SOC **(e)** the same with SOC.

**Fig. 8** The band dispersion in kz=0 plane (a) for without SOC (b) with SOC. The Dirac cone for without SOC is gapped out.

**Fig. 9(a)** The (110) plane is shown with the path marked for surface spectrum calculation. **(b)** edge states spectrum of PdSn$_4$ system along the path shown in 9(a). **(c)** Bulk spectral function plot of studied PdSn$_4$ system.

Fig. 1(a)

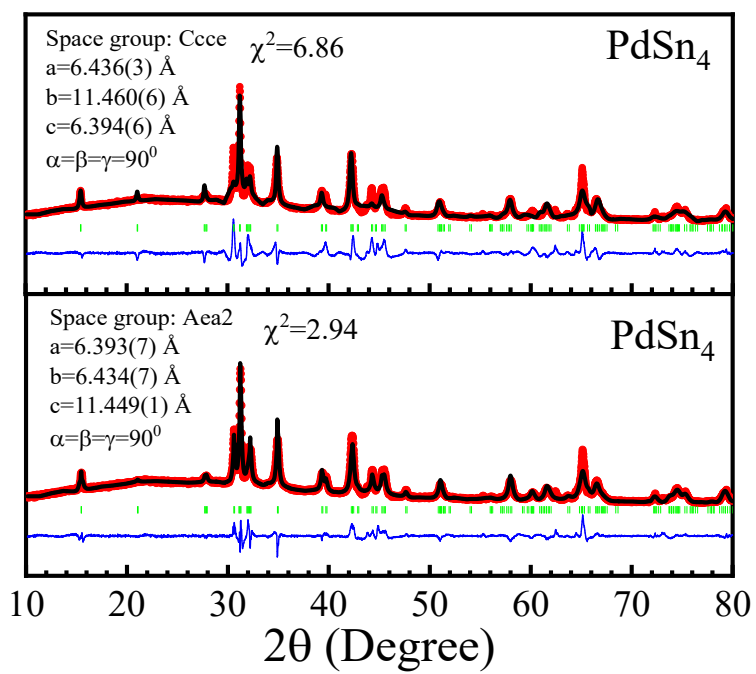

Fig. 1(b)&(c)

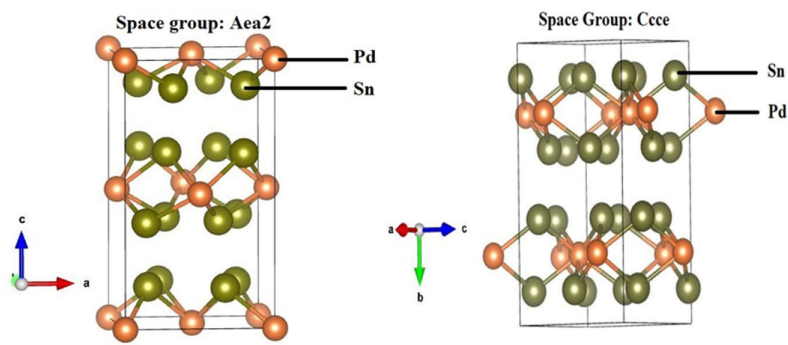

Fig. 1(d)

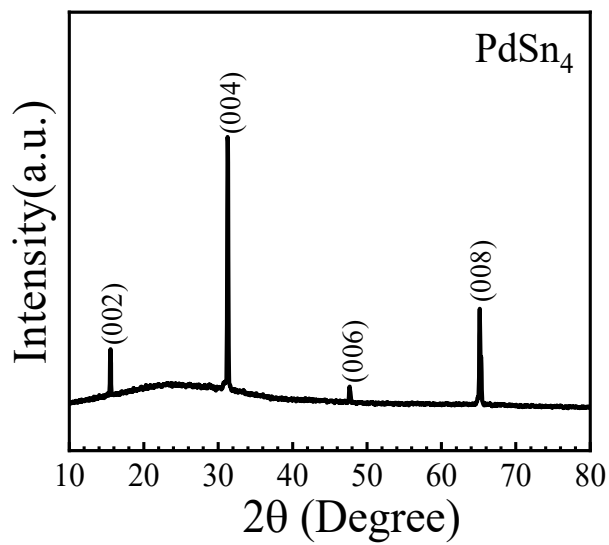

Fig. 2

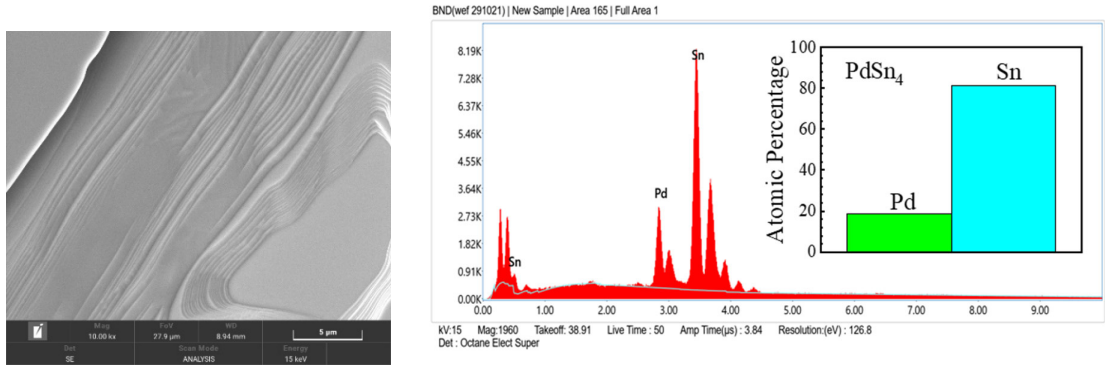

Fig. 3(a)&(b)

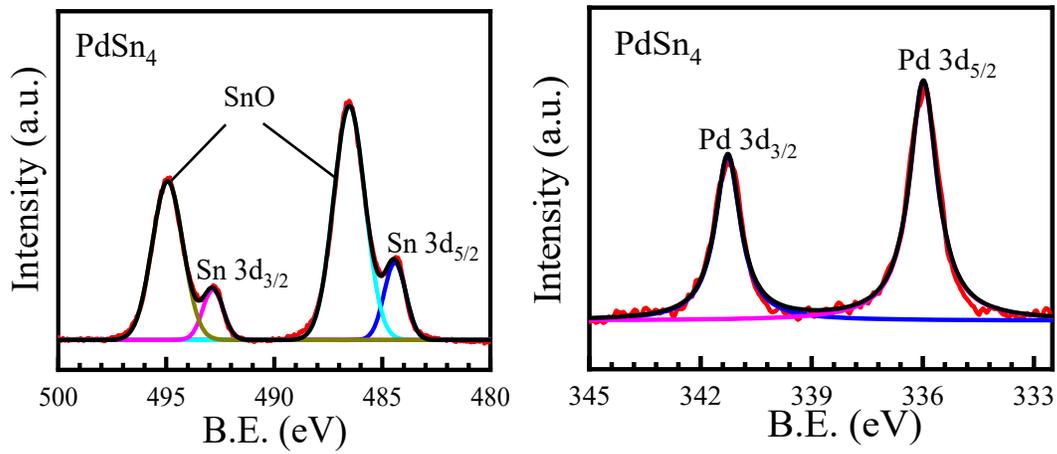

Fig. 4(a)

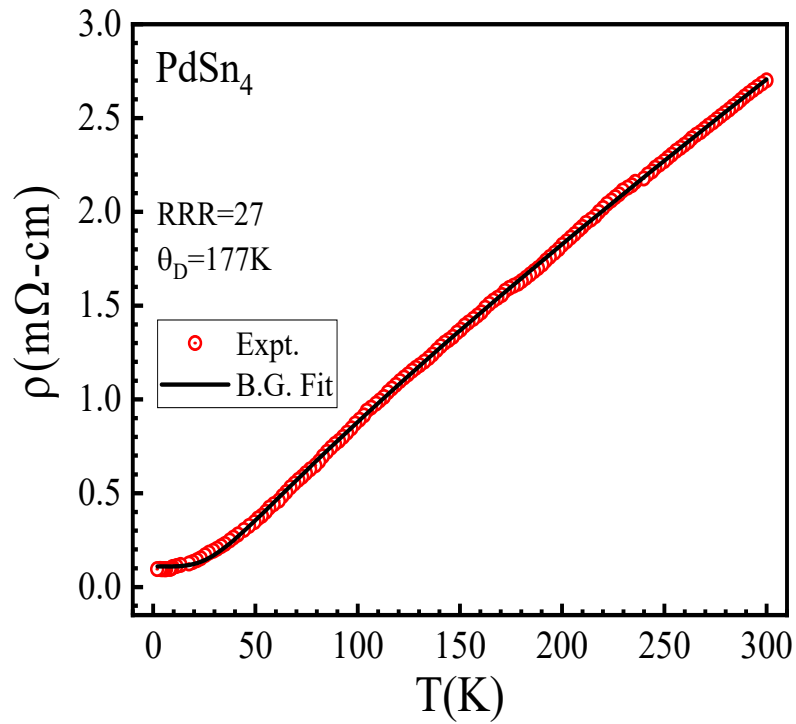

Fig. 4(b)

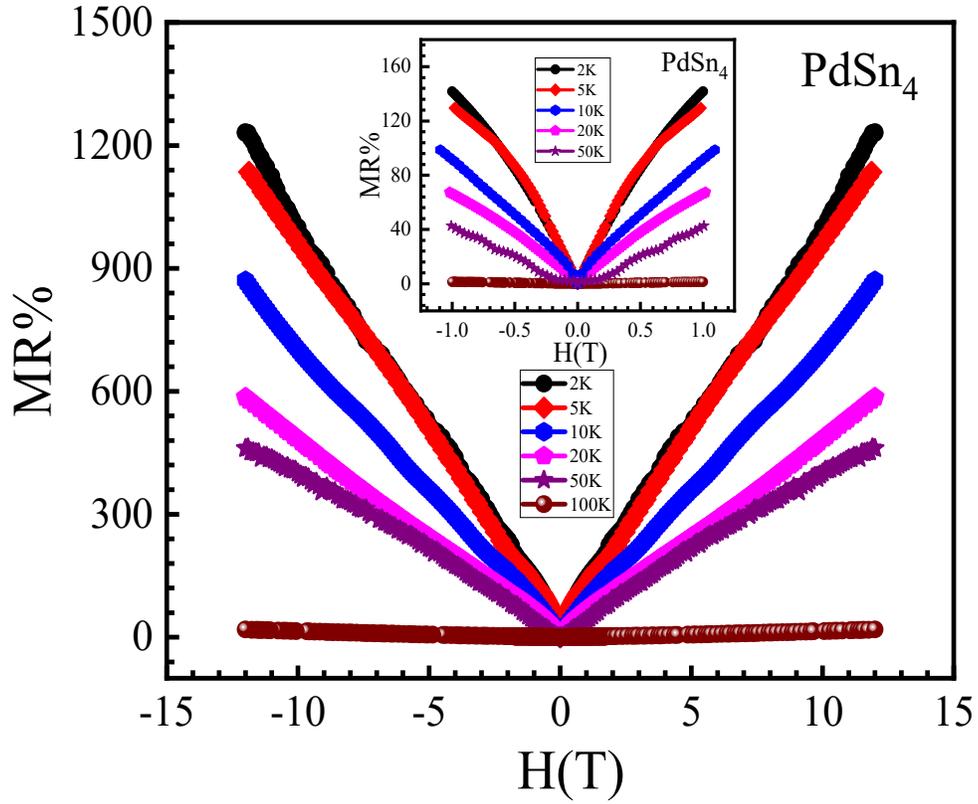

Fig. 4(c)

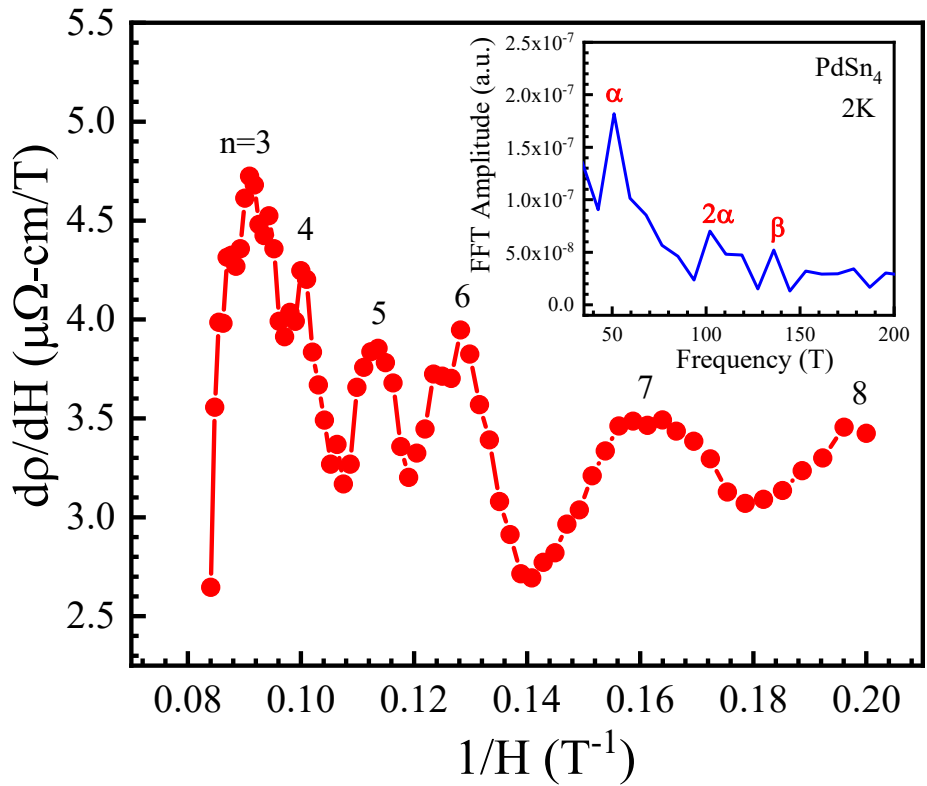

Fig. 4(d)

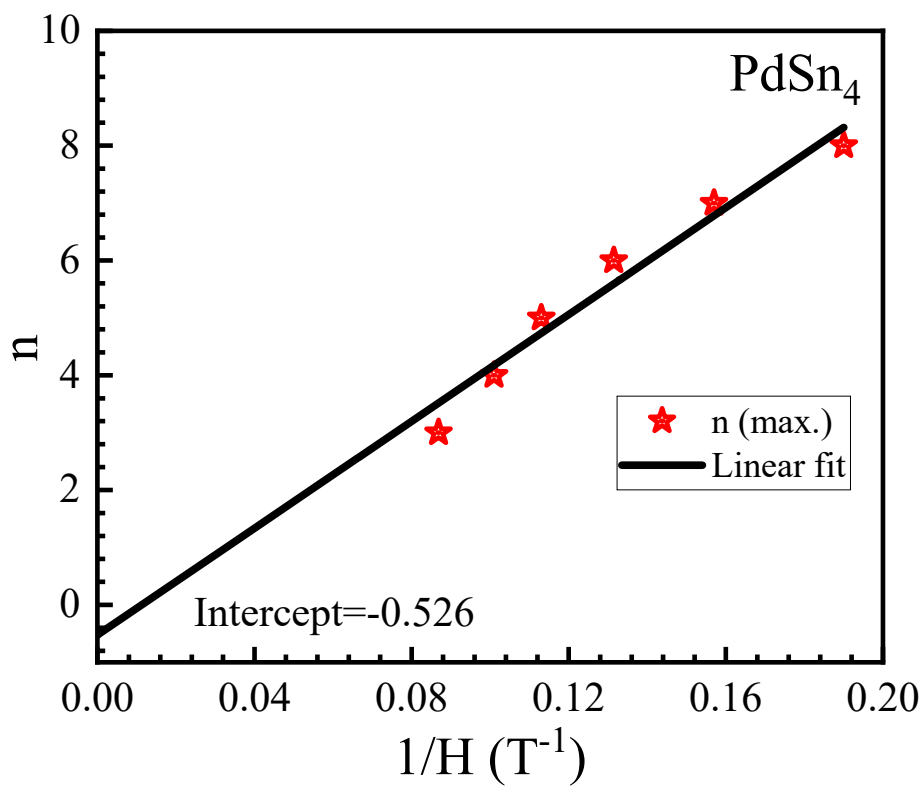

Fig. 5 (a)&(b)

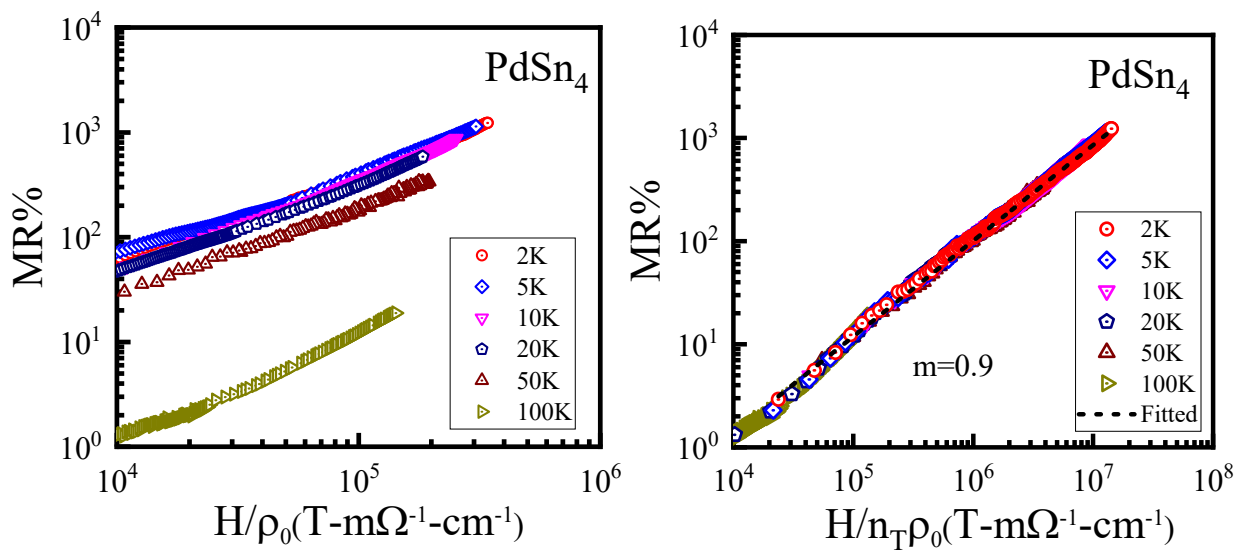

Fig. 6(a)

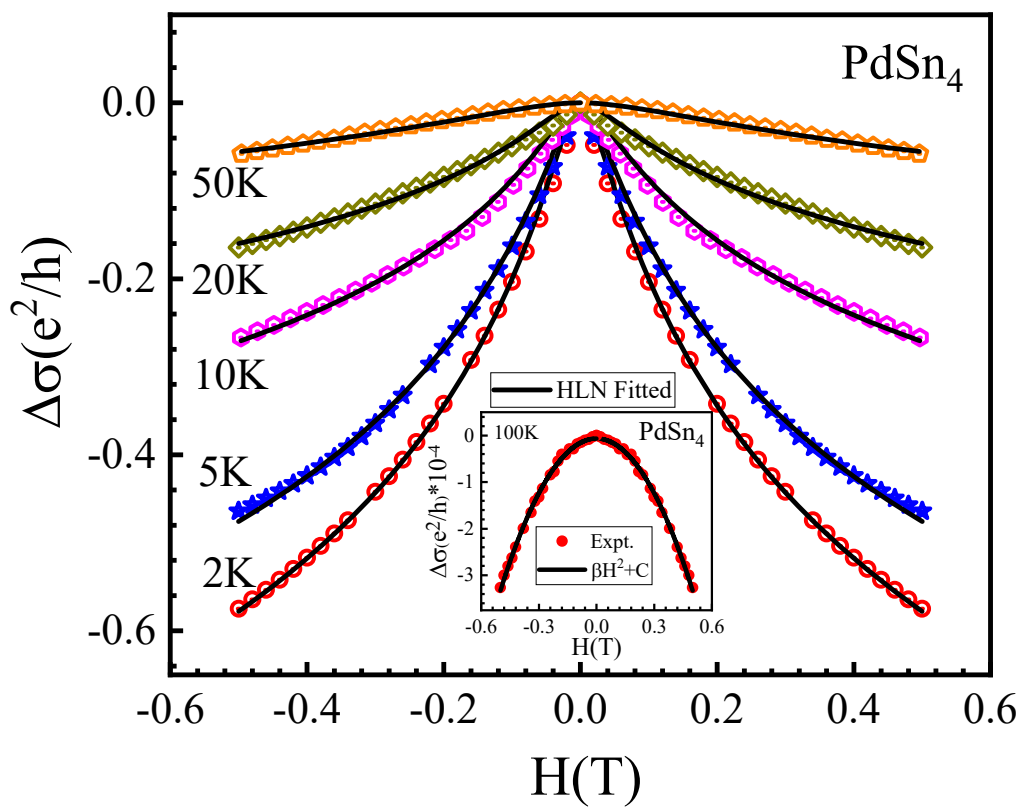

Fig. 6(b)

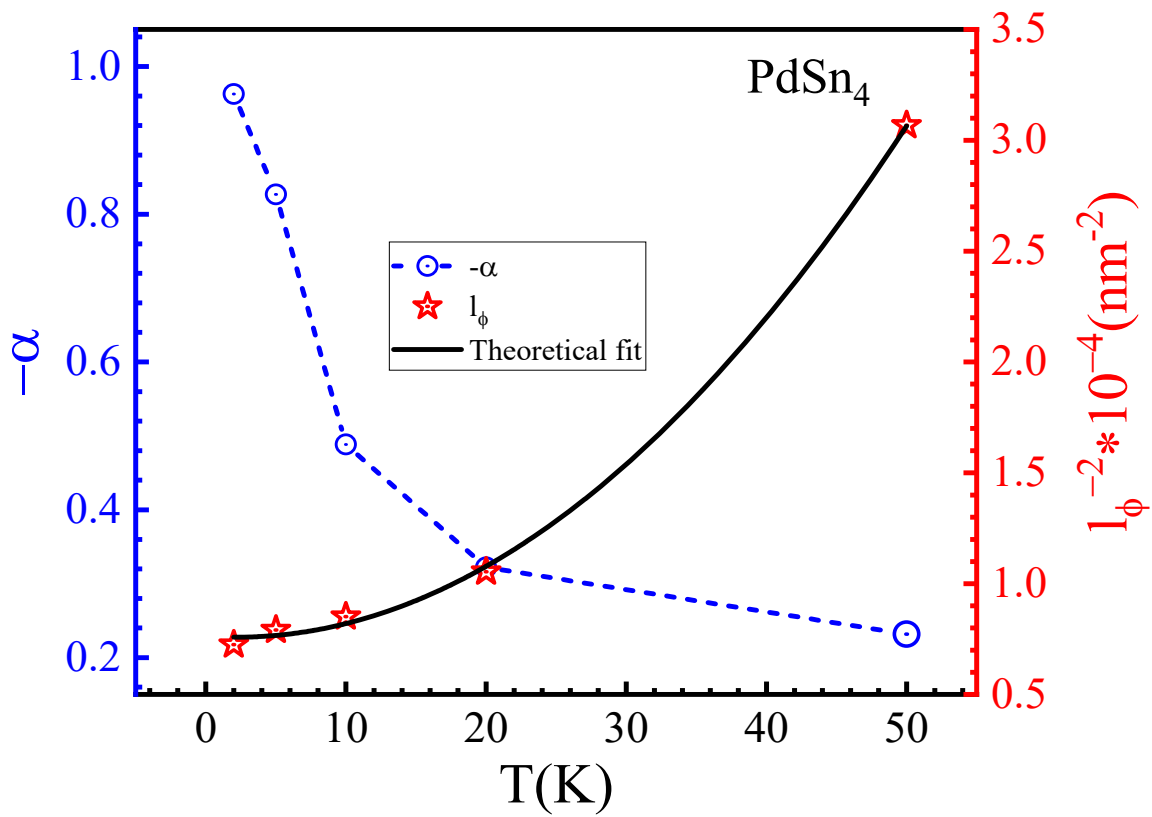

Fig. 7(a)

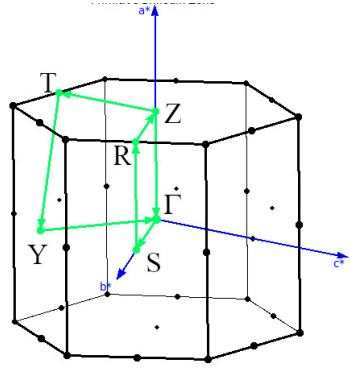

Fig. 7(b)

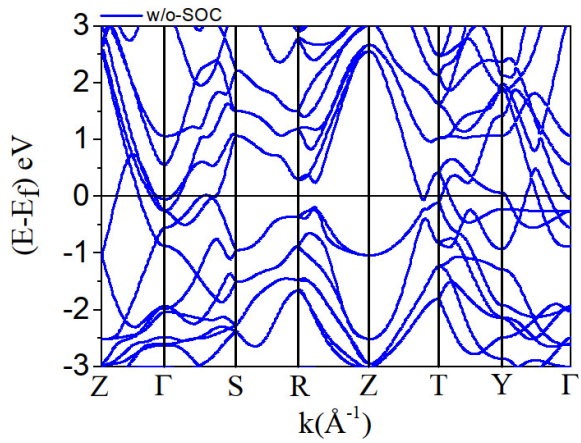

Fig. 7(c)

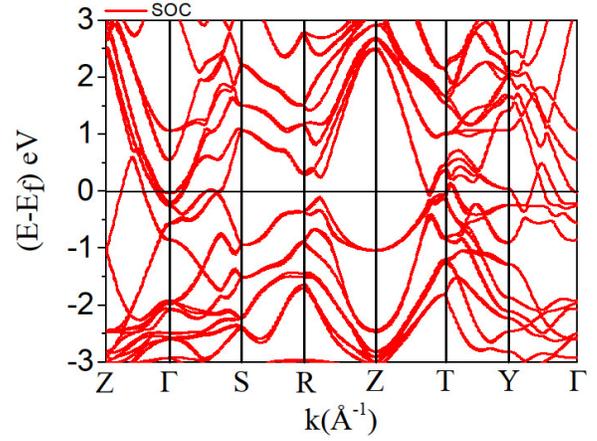

Fig. 7(d)

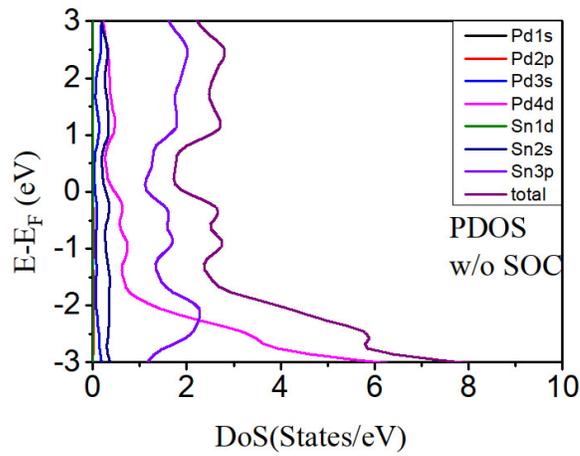

Fig. 7(e)

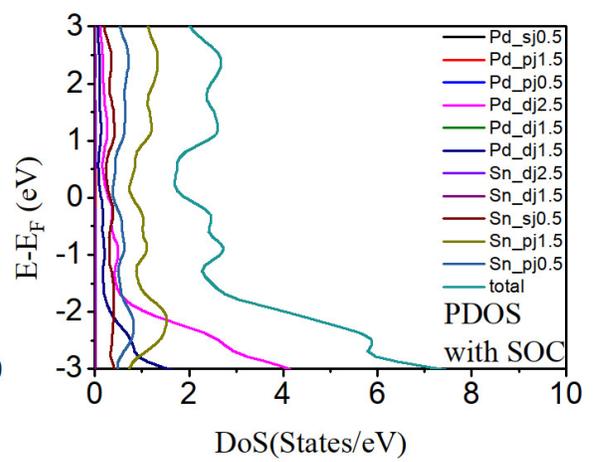

Fig. 8(a)

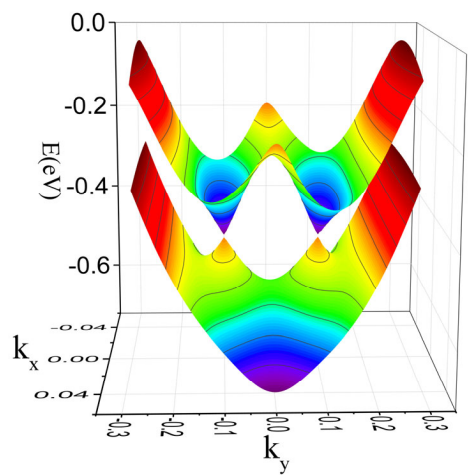

Fig. 8(b)

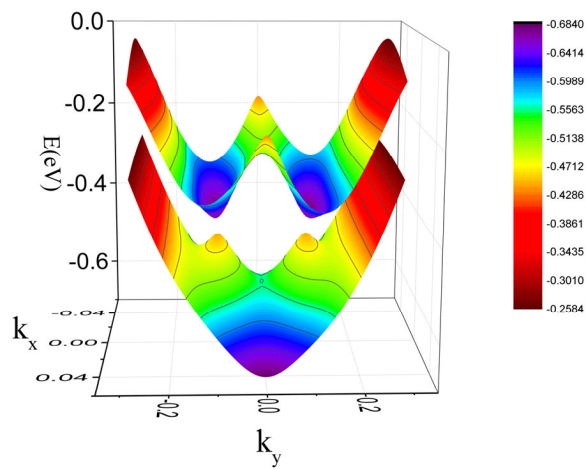

Fig. 9(a)

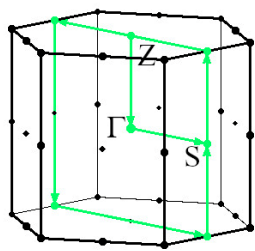

Fig. 9(b)

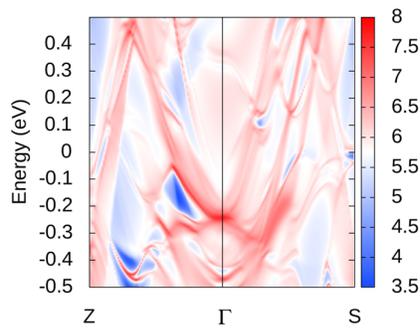

Fig. 9(c)

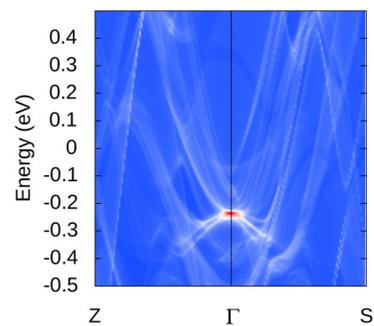